\documentclass[twocolumn,aps,superscriptaddress,showpacs, showkeys, nofootinbib,floatfix,linenumbers]{revtex4}
\usepackage{epsfig,bm,feynmf}
\usepackage{graphics}
\usepackage{amsmath}
\usepackage{mathrsfs}
\usepackage{appendix}
\usepackage{bm}
\usepackage[normalem]{ulem}  % \sout{old text} for strikeout
\usepackage[dvips]{color} % For blue in-text comments and additions

\renewcommand{\sout}{\bgroup \color{red} \ULdepth=-.5ex \ULset}

\begin{document}
\title{Three-particle correlations in relativistic heavy ion collisions in a multiphase transport model}	

% authors
\author{Yifeng Sun}
\email{sunyfphy@physics.tamu.edu}
\affiliation{Cyclotron Institute and Department of Physics and Astronomy, Texas A$\&$M University, College Station, Texas 77843, USA}%

\author{Che Ming Ko}
\email{ko@comp.tamu.edu}
\affiliation{Cyclotron Institute and Department of Physics and Astronomy, Texas A$\&$M University, College Station, Texas 77843, USA}%

% date
\date{\today}% It is always \today, today,

\begin{abstract}
Using a multiphase transport model, we study three-particle mixed harmonic correlations in relativistic heavy ion collisions by considering the observable $C_{m,n,m+n}=\langle\negmedspace\langle \cos(m\phi_1+n\phi_2-(m+n)\phi_3)\rangle\negmedspace\rangle$, where $\phi_{1,2,3}$ are azimuthal angles of all particle triplets. We find that except for $C_{123}$, our results on the centrality dependence of $C_{112}$, $C_{224}$ and $C_{235}$ as well as the relative pseudorapidity dependence of $C_{123}$ and $C_{224}$ in Au+Au collisions at $\sqrt{s}=$200 GeV agree reasonable well with the experimental data from the STAR Collaboration. We discuss the implications of our results.
\end{abstract}

\maketitle

\section{introduction}

The study of anisotropic flow in relativistic heavy ion collisions has provided important information on the properties of the produced quark-gluon plasma (QGP).  In earlier studies, the large elliptic flow observed in experiments at the BNL Relativistic Heavy Ion Collider (RHIC) for non-central collisions was found to be describable by ideal hydrodynamics. This has led to the conclusion that the produced QGP is an ideal fluid and thus a strongly interacting matter~\cite{Adams:2005dq,Adcox:2004mh,Arsene:2004fa,Back:2004je}.  More recent studies indicate that the experimental data on anisotropic flows could be better understood using viscous hydrodynamics ~\cite{Romatschke:2007mq,Song:2010mg,Schenke:2010rr} with a specific viscosity that is only about a factor of two larger than the theoretically predicted lower bound~\cite{Kovtun:2004de}.  In particular, the larger triangle flow observed in experiments not only put a more stringent constraint on the specific viscosity of the QGP~\cite{Gale:2012rq} but also reveal the importance of initial spatial fluctuations in heavy ion collisions~\cite{PhysRevC.81.054905}.  To study the
effect of initial-state fluctuations and that of final-state interactions with better precision, new observables based on $n$-particle correlations have been proposed~\cite{PhysRevC.84.034910}, since the ratio between these correlations and corresponding anisotropic flows can provide information on initial fluctuations. In the present study, we use a multiphase transport (AMPT) model~\cite{PhysRevC.72.064901} to study the three-particle correlations and compare the results with recent experimental measurements by the STAR Collaboration~\cite{Adamczyk:2017hdl,Adamczyk:2017byf}.

The paper is organized as follows. In the next section, we briefly describe the AMPT model and the parameters used in our calculations. In Sec. III, both two-particle and three-particle correlations are described.  Results on anisotropic flows and three-particle correlations obtained from the AMPT model are presented and compared with experimental data in Sec. IV. Finally, a summary is given in Sec. IV.

\section{The AMPT model}

The AMPT model is a hybrid model consisting of four stages of heavy ion collisions at ultrarelativistic energies: initial conditions, parton scatterings, conversion from the partonic matter to hadronic matter, and hadron scatterings~\cite{PhysRevC.72.064901} . There are two versions of the AMPT model, which are the default AMPT model and the AMPT model with string melting. In both versions, the initial conditions are generated from the heavy ion jet interaction generator (HIJING) model~\cite{PhysRevD.44.3501}. In the default version, only minijet partons from HIJING are included in the partonic stage via Zhang's parton cascade (ZPC)~\cite{ZHANG1998193}. After their scatterings, minijet partons are recombined with their parent strings to form excited strings, which are then converted to hadrons through the Lund string fragmentation model. In the string melting version, all hadrons produced from HIJING are converted to partons according to their valence quark flavors and spin structure, and these partons are evolved via the ZPC.  At the end of their scatterings, quarks and antiquarks are converted to hadrons via a simple coalescence model. Specifically, two nearest quark and antiquark are combined into a meson, and three nearest quarks (antiquarks) are combined into a baryon (antibaryon), with their species determined by  the flavor and invariant mass of coalescing quarks and antiquarks. Scatterings among hadrons in both the default and string melting versions are described by a relativistic transport (ART) model~\cite{PhysRevC.52.2037} until kinetic freeze-out.

In the present study, we use the string melting version of the AMPT model with the parameter set B of
Ref.~\cite{PhysRevC.84.014903}, i.e., using the values $a=0.5$ and $b=0.9$~GeV$^{-2}$ in the Lund string fragmentation function $f(z)\propto z^{-1}(1-z)^a\exp{(-bm_\perp^2/z)}$, where $z$ is the light-cone momentum fraction of the produced hadron of transverse mass $m_\perp$ with respect to that of the fragmenting string; and the values $\alpha_s=0.33$ and $\mu=3.2$ fm$^{-1}$ in the parton scattering cross section $\sigma\approx 9\pi\alpha_s^2/(2\mu^2)$.  This parameter set has been shown to give a better description of the charged particle multiplicity density, transverse momentum spectrum, and elliptic flow in heavy ion collisions at RHIC.

\section{Two- and three-particle correlations}

The two-particle correlation of particles in certain rapidity and transverse momentum range in heavy ion collisions is defined by~\cite{PhysRevC.83.044913}
\begin{eqnarray}\label{twoa}
c_n\{2\}&=&\left\langle\negmedspace\left\langle\frac{\sum_{i\ne j}\cos(n(\phi_i-\phi_j))}{M(M-1)}\right\rangle\negmedspace\right\rangle,
\end{eqnarray}
where the sum is over all possible particle pairs $i$ and $j$ in a single event with $M$ particles in that rapidity and transverse momentum range, $\phi_i$ and $\phi_j$ are azimuthal angles of their transverse momenta in the transverse plane of the collision, and the $\langle\langle\cdot\rangle\rangle$ denotes the average over events. Using the identity $\sum_{i\ne j}=\sum_{i,j}-\sum_{i=j}$, the numerator in the above equation can be written as
\begin{eqnarray}\label{twob}
&&\sum_{i\ne j}\cos(n(\phi_i-\phi_j))\nonumber\\
&&=M^2\left(\langle\cos n\phi\rangle^2+\langle\sin n\phi\rangle^2-\frac{1}{M}\right),
\end{eqnarray}
where $\langle\cdot\rangle$ denotes the average over all particles in a single event.  In the two-particle cumulant method~\cite{PhysRevC.83.044913}, anisotropic flow coefficients are simply given by the square root of the two-particle correlation, i.e., $v_n\{2\}=\sqrt{c_n\{2\}}$.

Similarly, the three-particle correlation, denoted as $C_{m,n,m+n}$, is defined by~\cite{PhysRevC.84.034910,Adamczyk:2017hdl,Adamczyk:2017byf}
\begin{eqnarray}\label{three}
&&C_{m,n,m+n}=\nonumber\\
&&\left\langle\negmedspace\left\langle\frac{\sum_{i\ne j\ne k}\cos(m\phi_i+n\phi_j-(m+n)\phi_k)}{M(M-1)(M-2)}\right \rangle\negmedspace\right\rangle.
\end{eqnarray}
Using the identity $\sum_{i\ne j \ne k}=\sum_{i,j,k}-\sum_{j=i,k}-\sum_{k=i,j}-\sum_{i,k=j}+2\sum_{i=j=k}$
~\cite{DiFrancesco:2016srj}, the numerator can also be written as
\begin{eqnarray}
&&\sum_{i\ne j\ne k}\cos(m\phi_i+n\phi_j-(m+n)\phi_k)=\nonumber\\
&&\sum_{i,j,k}\cos(m\phi_i+n\phi_j-(m+n)\phi_k)\nonumber\\
&&-\sum_{i,j}\cos(m(\phi_i-\phi_j))-\sum_{i,j}\cos(n(\phi_i-\phi_j))\nonumber\\
&&-\sum_{i,j}\cos((m+n)(\phi_i-\phi_j))+2M\nonumber\\
&&=M^3(\left \langle \cos m\phi \right \rangle\left \langle \cos n\phi \right \rangle\left \langle \cos(m+n)\phi \right \rangle\nonumber\\
&&-\left \langle \sin m\phi \right \rangle\left \langle \sin n\phi \right \rangle\left \langle \cos(m+n)\phi \right \rangle\nonumber\\
&&+\left \langle \sin m\phi \right \rangle\left \langle \cos n\phi \right \rangle\left \langle \sin(m+n)\phi \right \rangle\nonumber\\
&&+\left \langle \cos m\phi \right \rangle\left \langle \sin n\phi \right \rangle\left \langle \sin(m+n)\phi \right \rangle)\nonumber\\
&&-M^2(\left \langle \cos m\phi \right \rangle^2+\left \langle \sin m\phi \right \rangle^2)\nonumber\\
&&-M^2(\left \langle \cos n\phi \right \rangle^2+\left \langle \sin n\phi \right \rangle^2)\nonumber\\
&&-M^2(\left \langle \cos(m+n)\phi \right \rangle^2+\left \langle \sin(m+n)\phi \right \rangle^2)+2M,
\end{eqnarray}
which shows that the number of terms in the three-particle correlation can be reduced from $M^3$ to the order of $M$, thus improving significantly the efficiency of the calculation when $M$ is large.

\section{Results}

In this section, we show the anisotropic flow of charged particles and their three-particle mixed harmonic correlations obtained from the AMPT model in Au+Au collisions at $\sqrt{s_{\rm NN}}=200$ GeV at RHIC and compare them with experimental data measured by the STAR Collaboration.

\subsection{Charged particle anisotropic flow}

\begin{figure}[h]
\centering
\includegraphics[width=0.5\textwidth]{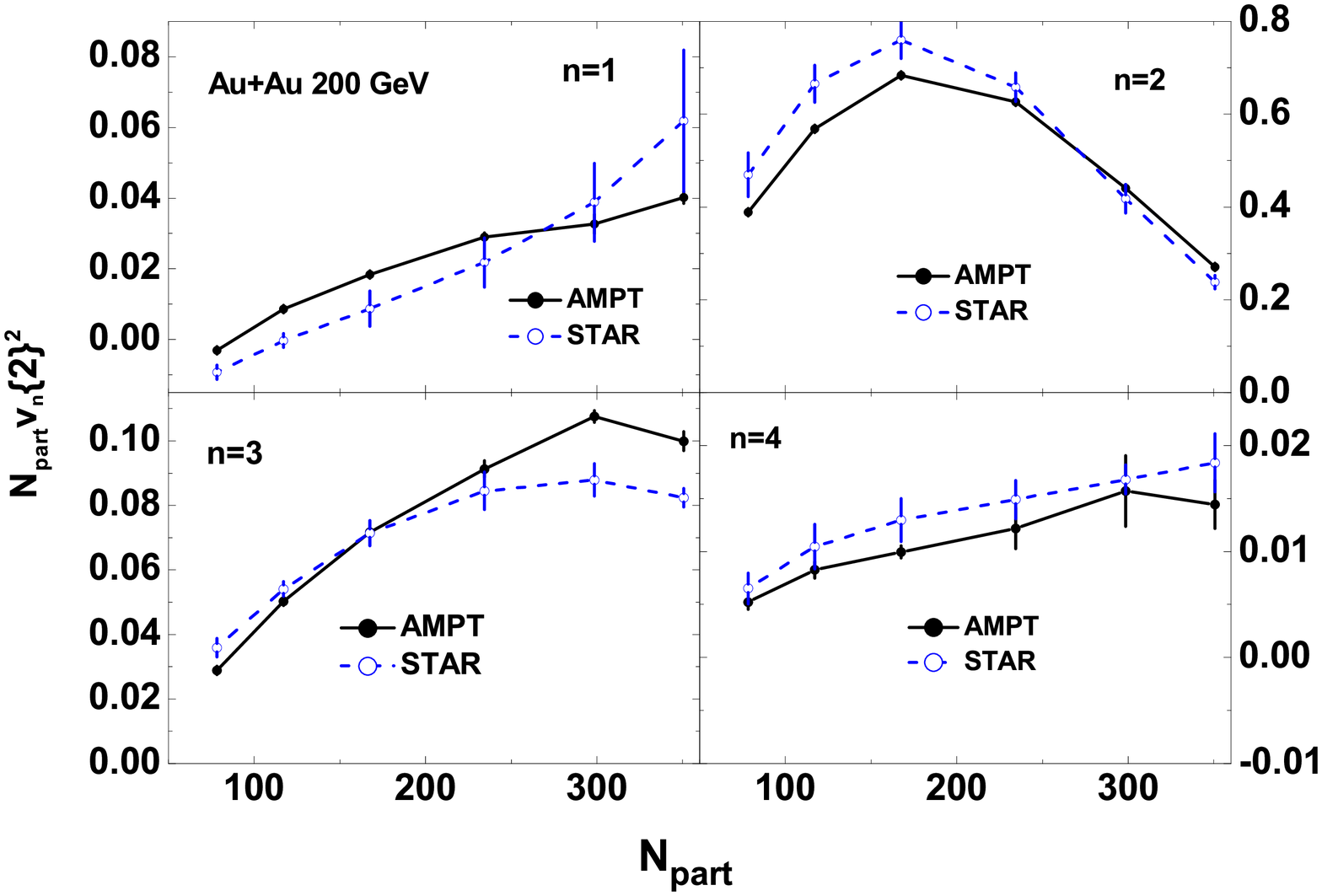}
\caption{(Color online) Participant number $N_{\rm part}$ or centrality dependence of $N_{\rm part}v_{n}\{2\}^2$ for mid-pseudorapidity ($|\eta|<1$) charged particles of transverse momentum $p_T>0.2$ GeV/$c$ in Au+Au collisions at $\sqrt{s_{\rm NN}}=200$ GeV. Open circles are experimental data from the STAR Collaboration~\cite{Adamczyk:2017hdl,PhysRevLett.116.112302}.}
\label{vn}
\end{figure}

In Fig. \ref{vn}, we show the participant number $N_{\rm part}$ or centrality dependence of anisotropic flow from $n=1$ to 4 for mid-pseudorapidity ($|\eta|<1$) charged particles of transverse momentum $p_T>0.2$ GeV/$c$ in Au+Au collisions at $\sqrt{s_{\rm NN}}=200$ GeV.  In particular, the participant numbers chosen in our calculations are those in collisions at impact parameters of $2.2$, $4.1$, $5.8$, $7.5$, $8.8$, and $10.0$ fm, corresponding, respectively, to centrality bins of 0-5\%, 5-10\%, 10-20\%, 20-30\%, 30-40\% and 40-50\% in the STAR experiment.  We calculate the anisotropic flow from the two-particle cumulant method~\cite{PhysRevC.83.044913}.  Note that we do not include the anisotropic flow for $n=5$ because of the large uncertainty in both experimental data~\cite{Adamczyk:2017byf} and our results.  It is seen that the results from the AMPT model agree qualitatively with the experimental data~\cite{Adamczyk:2017hdl,PhysRevLett.116.112302}. Quantitatively, the AMPT slightly overestimates the measured elliptic flow $v_2\{2\}$ and triangular flow $v_3\{2\}$ for the most-central collisions and underestimate them for peripheral collisions.  On the other hand, our results for the directed flow $v_1\{2\}$ underestimate for the most-central collisions and overestimate for peripheral collisions. For the quadrupolar flow, our results are slightly smaller than the data for all centralities.

In both experimental data and results from the AMPT, $v_1\{2\}^2$ are negative for more peripheral collisions.  This can be understood from Eq.(\ref{twob}). In the case of including all particles in a collision, the first term $\langle\cos\phi\rangle^2+\langle\sin\phi\rangle^2$ should be identically zero due to the conservation of total transverse momentum. Since only mid-pseudorapidity ($|\eta|<1$) particles of transverse momentum $p_T>0.2$ GeV/$c$ are included in calculating $v_1\{2\}^2$,
the value of the first term can be nonzero but small.  This can lead to positive values of $v_1\{2\}^2$ for large $M$. With decreasing particle number $M$ for more peripheral collisions, the first term in Eq.(\ref{twob}) can become smaller than the second term $1/M$, resulting in a negative value for $v_1\{2\}^2$.

\subsection{Three-particle correlations}

\begin{figure}[h]
\centering
\includegraphics[width=0.5\textwidth]{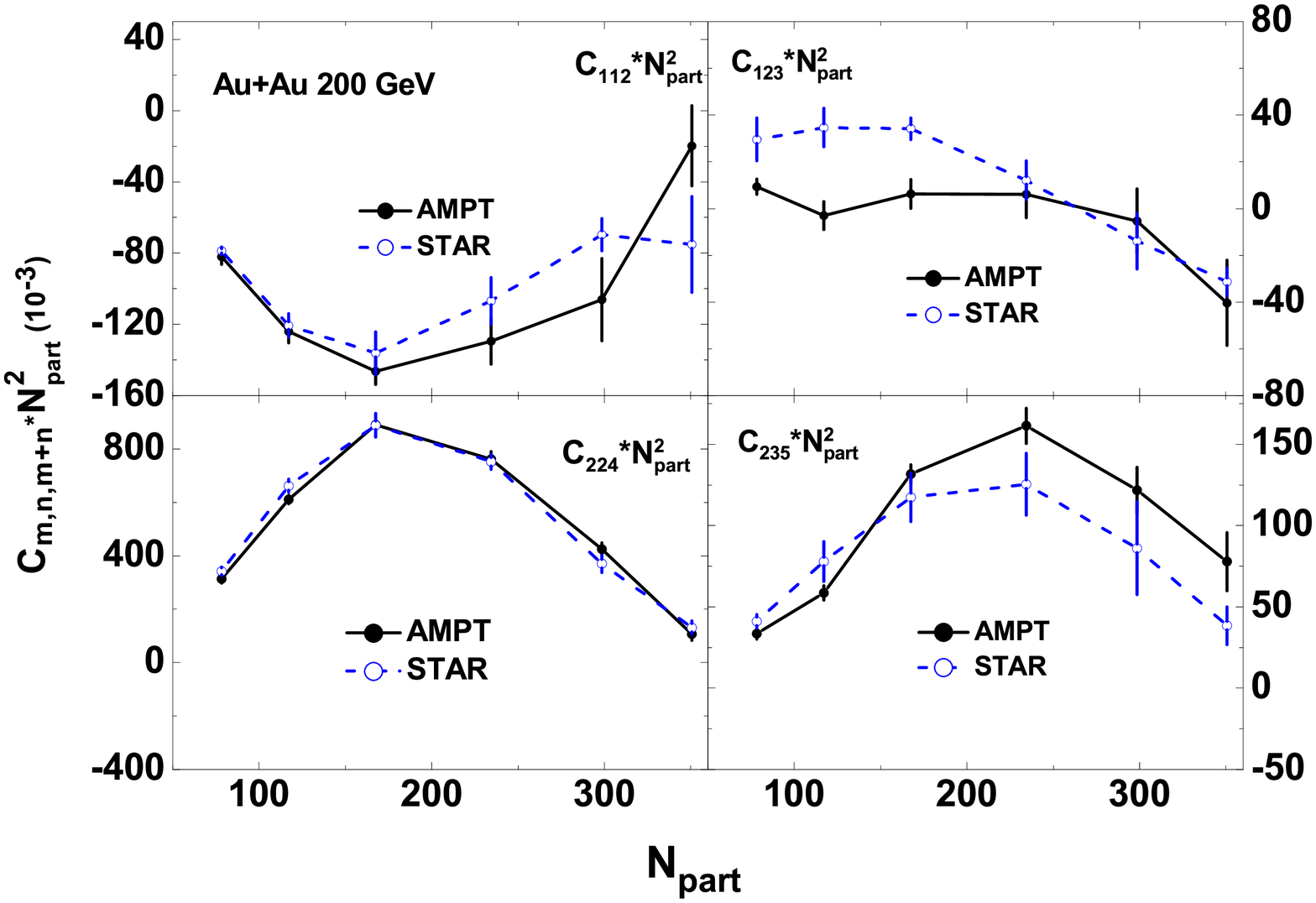}
\caption{(Color online) Centrality dependence of $C_{m,n,m+n}\times N_{\rm part}^2$ for  mid-pseudorapidity ($|\eta|<1$) charged particles of transverse momentum $p_T>0.2$ GeV/$c$ in Au+Au collisions at $\sqrt{s_{\rm NN}}=200$ GeV.  Open circles are experimental data from Ref.~\cite{Adamczyk:2017byf}.}
\label{corr}
\end{figure}

For the three-particle correlations, we have calculated $C_{112}$, $C_{123}$, $C_{224}$ and $C_{235}$, and compared them to the experimental results~\cite{Adamczyk:2017byf}. Figure \ref{corr} shows $C_{m,n,m+n}\times N_{\rm part}^2$ for the four cases as functions of the number of participant nucleons.  For $C_{112}$ in the upper left panel of Fig.~\ref{corr}, our results show good agreement with the experimental data, although there are some discrepancy in more central collisions.  Values of $C_{112}$ from both our calculations and the experimental measurement are negative for all centralities.  Besides possible non-flow effects from momentum conservation in the AMPT simulations, this could imply that the angles $\Psi_1$ and $\Psi_2$ of the reaction planes for directed and elliptic flows are likely to be perpendicular to each other.  Our results on $C_{123}$, shown in the upper right panel of Fig. \ref{corr}, are seen to agree with the experimental data within their error bars for most-central collisions but are smaller than the experimental data for mid-central collisions. Their essentially zero values indicate that the directed and triangular flows or the angles $\Psi_1$ and $\Psi_3$ of their reaction planes are not sufficiently correlated in the AMPT model for mid-central collisions.  For $C_{224}$ shown in the lower left panel, our results agree extremely well with experimental data for all centralities, although there are small difference between our results on elliptic and quadrupolar flows and those measured in experiments as shown in Fig.~\ref{vn}.  Their large values further indicate that there is a strong correlation between the angles $\Psi_2$ and $\Psi_4$ of their reaction planes.  The lower right panel shows our results for $C_{235}$, which are seen to show similar trend and magnitude as experimental data, although overestimating the data in most-central collisions and underestimating it in mid-central collisions. Since the AMPT reproduces reasonably well various anisotropic flows measured in experiments as shown in Fig.~\ref{vn}, the above results on its reasonable success in describing also the measured three-particle correlations clearly indicate that the initial states in the AMPT model are quite similar to what are generated in heavy ion collisions.

\subsection{Relative pseudorapidity dependence of $C_{m,n,m+n}$}

\begin{figure}[h]
\raggedleft
\includegraphics[width=0.5\textwidth]{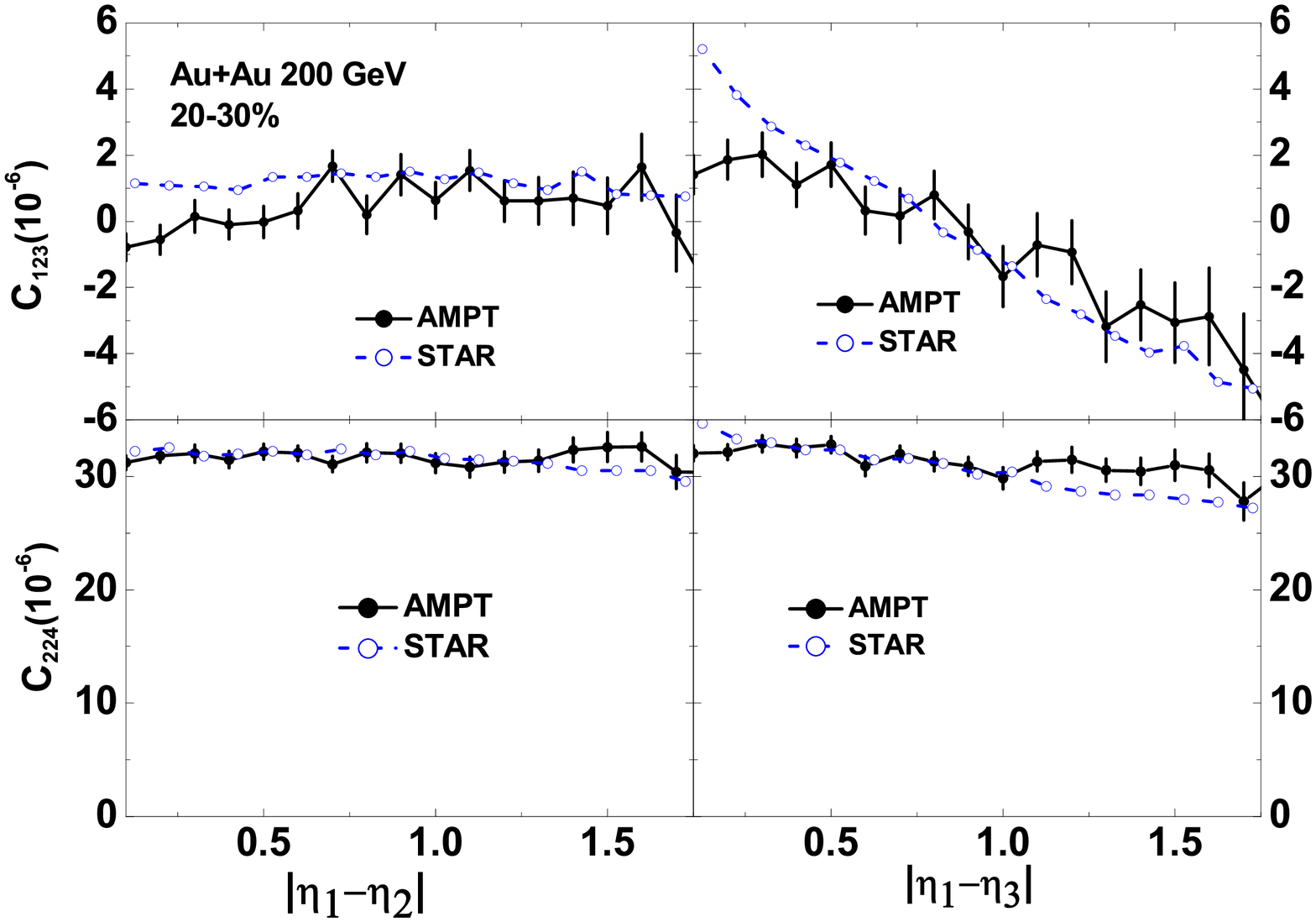}
\caption{(Color online) Relative pseudorapidity $|\Delta \eta|$ dependence of $C_{m,n,m+n}$ for mid-pseudorapidity ($|\eta|<1$) charged particles of transverse momentum $p_T>0.2$ GeV/$c$ in Au+Au collisions at $\sqrt{s_{\rm NN}}=200$ GeV. Open circles are experimental data from Ref.~\cite{Adamczyk:2017byf}.}
\label{eta}
\end{figure}

In this section, we study the relative pseudorapidity $|\Delta \eta|$ dependence of $C_{m,n,m+n}$ for mid-pseudorapidity ($|\eta|<1$) charged particle of transverse momentum $p_T>0.2$ GeV/$c$ in Au+Au collisions at $\sqrt{s_{NN}}=200$ GeV and centrality 20-30\%. In particular, we consider the pseudorapidity difference between the first and the second particle ($\eta_1-\eta_2$) or the first and the third particle ($\eta_1-\eta_3)$. The upper left panel of Fig. \ref{eta} shows that $C_{123}$ from the AMPT model only changes slightly with $|\eta_1-\eta_2|$ as in the experimental data. These results imply that there is negligible breaking of boost invariance as seen in terms of the pseudorapidity dependence of the angle $\Psi_2$ of elliptic flow. The $|\eta_1-\eta_3|$ dependence of the results from the AMPT model, shown in the upper right panel, shows, on the other hand, a strong decrease with increasing $|\eta_1-\eta_3|$ as in the data, indicating that azimuthal angles $\Psi_1$ and $\Psi_3$ of the reaction planes for the directed and triangular flows change strongly with rapidity. Since our results for $C_{123}$ are smaller than experimental data for small values of $|\eta_1-\eta_2|$ and $|\eta_1-\eta_3|$, the reaction planes for the directed and triangular flows in the AMPT model is thus less correlated than measured in experiments. This is probably due to the Hambury-Brown-Twiss interference of identical particles at small $\Delta\eta$~\cite{HanburyBrown:1956bqd}, which is not included in the AMPT.

The lower two panels show the $|\eta_1-\eta_2|$ and $|\eta_1-\eta_3|$ dependence of $C_{224}$, and both are seen to agree with experimental data very well.  As for $C_{123}$ in the upper left panel, $C_{224}$ in the lower left panel also changes little with $|\eta_1-\eta_2|$, indicating that the reaction plane for the elliptic flow has a weak dependence on rapidity.  The lower right panel shows that $C_{224}$ decreases slightly with increasing $|\eta_1-\eta_3|$, implying that the reaction plane for the quadrupolar flow changes with rapidity and thus breaks slightly the boost invariance.

\section{Summary}

Using the AMPT model with parameters for the Lund string fragmentation and parton scattering taken from Ref.~\cite{PhysRevC.84.014903}, we have calculated the centrality dependence of various anisotropic flows in Au+Au collisions at $\sqrt{s_{\rm NN}}=200$ GeV from the two-particle cumulant method.  The obtained results are seen to agree with experimental data from the STAR Collaboration in both the trend and magnitude. We have found that the square of the directed flow $v_{1}\{2\}^2$ can be negative in more peripheral collisions as in experiments, and this has been attributed to the net total transverse momentum of particles included in the evaluation and the small number of particles in more peripheral collisions.

We have also used the AMPT model to study various three-particle correlations in Au+Au collisions at $\sqrt{s_{\rm NN}}=200$ GeV as functions of centrality, which contain information on both flow harmonics and correlations among their reaction planes. We have found that our results for $C_{112}$, $C_{224}$ and $C_{235}$ generally agree with experimental data both in their magnitude and dependence on the participant number of collisions. In particular, our results for $C_{224}$ agree very well with the data, although our results for the elliptic and quadrupolar flows differ slightly from the data.  For $C_{123}$, our results show that for mid-central collisions there is a weaker correlation between the angles of the reaction plane for the directed, elliptic and triangular flows for mid-central collisions in AMPT model than in the experimental data. We have further studied the dependence of three-particle correlations on the relative pseudorapidity $|\eta_1-\eta_2|$ and $|\eta_1-\eta_3|$ between first and second particles as well as between first and third particles. Our results are seen to agree with experimental data for $C_{123}$ and $C_{224}$, and indicate that the boost invariance is weakly broken in the angles of the reaction planes for the elliptic and quadrupolar flows but strongly broken in those for the directed and triangular flows.  These results have led us to conclude that the AMPT model with its fluctuating initial conditions and strong partonic scatterings can capture the essential collision dynamics of relativistic heavy ion collisions as revealed in the measured anisotropic flows and three-particles correlations.

\section*{ACKNOWLEDGEMENTS}

We thank Prithwish Tribedy for discussions that led to present study and for his critical reading of the manuscript. This work was supported in part by the US Department of Energy under Contract No. DE-SC0015266 and the Welch Foundation under Grant No. A-1358.

\bibliography{ref}
\end{document}